\documentclass[twoside,reqno]{qts-proc}
\usepackage{epsfig,cite}
\usepackage{amssymb,amsmath}
\usepackage{times}\usepackage{dsfont}
\newcommand{\ei}{\end{itemize}}
\newcommand{\be}{\begin{eqnarray}}
\newcommand{\ee}{\end{eqnarray}}
\newcommand{\ba}{\begin{array}}
\newcommand{\ea}{\end{array}}
\newcommand{\bc}{\begin{center}}
\newcommand{\ec}{\end{center}}
\newcommand{\bt}{\begin{tabular}}
\newcommand{\btab}{\begin{table}}
\newcommand{\et}{\end{tabular}}

\begin{document}
\sloppy \raggedbottom
\setcounter{page}{1}
\newpage
\setcounter{figure}{0}
\setcounter{equation}{0}
\setcounter{footnote}{0}
\setcounter{table}{0}
\setcounter{section}{0}
\title{Extension of Monster Moonshine to $c=24\, k$ Conformal Field Theories
\thanks{Lecture given by M. Jankiewicz at QTS-4, 15-21 Aug 2005, Varna Free University, Bulgaria.}}



\vspace{-0.7cm}
\author{Marcin M. Jankiewicz}{} and
\coauthor{Thomas W. Kephart}{},

\address{Department of Physics and Astronomy, Vanderbilt University, Nashville, TN 37235, USA}{}

\begin{Abstract}
\noindent
We present a family of conformal field theories (or candidates for CFTs) that is 
build on extremal partition functions. Spectra of these theories can be decomposed 
into the irreducible representations of the Fischer-Griess Monster sporadic group. 
Interesting periodicities in the coefficients of extremal partition functions are 
observed and interpreted as a possible extension of Monster moonshine to $c=24\,k$ 
holomorphic field theories.
\end{Abstract}


\section{Introduction}

We will construct partition functions of conformal field theories with central 
charge which is a multiple of $24$. Our construction is based on a unique modular function, 
the so called $j$-invariant (or hauptmodule). The properties of this function  guarantee (at least at modular level) 
that any integer power of it will be again a modular function and this is of key importance in our construction.\\
Moreover, we report on a possible extension of a Monster moonshine, that relates 
coefficients in a $q$-expansion of $j$-invariant function and dimensions of the irreducible
 representations of the Monster sporadic group.
We are going to use lattices, or more precisely their $\Theta$-functions to describe a given CFT. 
The $q$-expansion of a $\Theta$-function of a lattice $\Lambda$ is given as
\be Z_{\Lambda}=\sum_{x\in\Lambda}N(m)q^{m}\,,\ee
where we sum over all vectors $x$, in the lattice $\Lambda$, with a length $m=x\cdot x$. $N(m)$ 
is the number of vectors of norm $m$ and $q\equiv e^{i\pi\tau}$ in terms of the modular parameter $\tau$.
The spectra of meromorphic conformal field theories can be expressed in terms of partition functions of even self-dual lattices.
This means that the exponent $m$ in the $q$-expansion will be necessarily an even number. 
Both self-duality and evenness of a lattice correspond to invariance of a partition function $\mathcal{Z}$, 
closely related to $Z_{\Lambda}$, under the generators $S$ and $T$ of a modular group $SL(2,\mathds{Z})$.\\
Formally, $Z_{\Lambda}$ of a $d$ dimensional lattice $\Lambda$ is a modular form of weight $d/2$. A
partition function $\mathcal{Z}$ of a lattice is defined as follows
\be\label{partdef}\mathcal{Z}=Z_{\Lambda}/\eta^{d/2}\,,\ee 
where $\eta(q)=q^{1/12}\prod_{m=1}^{\infty}(1-q^{2m})$
is Dedekind $\eta$-function which is a modular form of weight $1/2$.
Partition functions of a all the 24 dimensional even self-dual lattices (the Niemeier lattices) 
can be written as
\be\label{part1}\mathcal{Z}=\left[J+24(h+1)\right]\eta^{24}\,,\ee 
where $h$ is the Coxeter number of a given lattice. For example $h=0$ corresponds to the famous 
Leech lattice, $h=30$ to the Niemeier lattice based on a root system of $E_{8}^{3}$, etc. 
Physically $24(h+1)$ corresponds to a number of massless states in a given theory.\\ 
Using a technique presented in \cite{Jankiewicz:2005rx} one can choose any lattice $\Lambda_{1}$ 
to generate the $\Theta$-function of another lattice $\Lambda_{2}$. In the same paper it was shown 
that it is possible to generate a partition function of any extremal lattice (by extremal, 
we mean the lattice that has the tightest packing in a given dimension\footnote{If such a lattice exists.}) 
by taking the $k^{th}$ power of (\ref{part1}) and treating $x_{i}=24(h_{i}+1)$ (where $i=1,...,k$) as a free parameter.


\section{CFTs with $c=24k$: Systematic Approach}
Different choices of constant parameters $x_{i}s$ correspond to 
different $\Theta$-functions. In principle one can use the same technique to find corresponding extremal 
partition functions, that are related by (\ref{partdef}). One can write them as $q$-expansions of the form
\be\prod_{i=1}^{k}(J+24+x_{i})=\frac{1}{q^{2k}}\left[1+\sum_{m=(k-1)}^{\infty}f_{2m}(x_{1},...,x_{k})q^{2m-2k}\right]\ee
 Here we want to focus on two choices of these parameters that lead to interesting families of partition functions, 
that are motivated by both physics and mathematics.\\ 
The first choice is reminescent of the one introduced in the examples presented in the previous section, 
namely the choice of $k-1$ parameters $x_{i}$s such that the lattice has densest possible packing in a given dimension. 
More precisely, thanks to this parametrization, one eliminates coefficients of terms with negative powers in the 
$q$-expansion that correspond (in a field theoretic language) to tachyonic states. In this setup we are left with only one free parameter $x_{k}$. 
Different choices of $x_{k}$ would correspond to different partition functions of candidates for conformal field 
theories with $c=24\cdot k$. Here we list the first three cases:
\begin{subequations}
\begin{align}
&\mathcal{G}_{1}(x_{1})\!=\!\frac{1}{q^{2}}+(24+x_{1})+196884q^{2}+...\\
&\mathcal{G}_{2}(x_{2})\!=\!\frac{1}{q^{4}}+(393192-48x_{2}-x_{2}^{2})+42987520q^{2}+...\\ \label{tach}
&\mathcal{G}_{3}(x_{3})\!=\!\frac{1}{q^{6}}+(50319456-588924x_{3}+72x_{3}^{2}+x_{3}^{3})+2592899910q^{2}+...
\end{align}\end{subequations}
Notice that in each case all of the tachyonic states (except the lowest one) are absent. Since the allowed 
values \cite{Harvey:1988ur} of the coefficient of $q^{0}$ are integers that run from zero to the value of the 
$q^{2}$ coefficient, one can easily find the number of ``allowed'' partition functions in $24\cdot k$ dimensions.\\
There exists an interesting alternative $x$-parametrization \cite{Apostol} of the partition functions
\begin{subequations}
\begin{align}
&\mathcal{H}_{1}=\frac{1}{q^{2}}+196884q^{2}+...\\
&\mathcal{H}_{2}=\frac{1}{q^{4}}+1+42987520q^{2}+...\\
&\mathcal{H}_{3}=\frac{1}{q^{6}}+\frac{1}{q^{2}}+1+2593096794q^{2}+...
\end{align}
\end{subequations}
Here we fix the tachyon levels, i.e., the levels with $q^{m}$ where $m<0$, and the massless level, i.e., $q^{0}$, 
by appropriate choices of the $x$s, so that each level except the $(k-1)$th one contains a single state.
This parameterization is interesting since the first nontrivial coefficient corresponds to the characters of the 
extremal vertex operator algebra of rank $24\cdot k$.

\section{Monster Moonshine and its Extension}
The extremal 24 dimensional case has been shown to be related to the Fischer-Griess monster group. 
In mathematics this fact is known as Monster moonshine (\cite{Dolan:1989kf} and \cite{Borch}). 
One can evaluate $\mathcal{G}_{1}$ at $x_{1}=-24$ which corresponds to the $j$-invariant to find
\begin{eqnarray}j=\frac{1}{q^{2}}+196884q^{2}+ 21493760q^{4} + 864299970q^{6} + 20245856256q^{8}+...\,.\end{eqnarray}
\noindent
The coefficients of this expansion decompose into dimensions of the irreducible representations of the 
Monster\footnote{for explicit realization of the Monster moonshine see \cite{Jankiewicz:2005rx}.}, where we use the notation $j=\frac{1}{q^{2}}+j_{2}q^{2}+j_{4}q^{4}+...\,$.
Following this interpretation of the Monster moonshine theorem, one can easily generalize it to 
 higher dimensional cases, i.e., one can express coefficients of any partition functions, 
for example $\mathcal{G}_{k}(x_{k})$ or $\mathcal{H}_{k}$, for any choice of $k$, in terms of the dimensions of 
irreducible representations of the Monster group. We present the results in Table-\ref{tab-mon3}, 
where coefficients of both $G_{k}(x_{k})$ and $\mathcal{H}_{k}$ are expressed in terms of the coefficients of the invariant function $j$, 
that (via the original Monster moonshine) are related to the Monster.
We notice that the coefficients the $g_{2n}$ and $h_{2n}$ fall into patterns with period $k!$.
We conjecture that this periodicity also continues to hold for all $k$.
The polynomial conditions to be satisfied to find the
extremal partition functions for large $k$ become increasingly more
difficult to solve with increasing $k$, so we do not have results
for $k>6$.\\ 
Table-\ref{tab-mon3} give the general periodicity in coefficients $g_{2n}$ and $h_{2n}$ of, respectively, $\mathcal{G}_{k}(x_{k})$ and $\mathcal{H}_{k}$. 
\begin{table}\label{tab-mon3}

{\scriptsize\begin{tabular}{|l|l||l|l|}
\hline
$k=2$        & $k=2$                               & $k=2$     & $k=2$     \\ \hline
$g_{4i+2}$   & $2j_{2(4i+2)}$                      & $h_{4i+2}$& $2j_{2(4i+2)}$  \\
$g_{4i+4}$   & $2j_{2(4i+4)}+j_{2(2i+2)}$          & $h_{4i+4}$& $2j_{2(4i+4)}+j_{2(2i+2)}$  \\ \hline

$k=3$        & $k=3$                               & $k=3$     & $k=3$  \\ \hline
$g_{6i+2}$   & $3j_{3(6i+2)}$                      & $h_{6i+2}$& $3j_{3(6i+2)}+j_{6i+2}$  \\
$g_{6i+4}$   & $3j_{3(6i+4)}$                      & $h_{6i+4}$& $3j_{3(6i+4)}+j_{6i+4}$  \\
$g_{6i+6}$   & $3j_{3(6i+6)}+j_{2i+2}$             & $h_{6i+6}$& $3j_{3(6i+6)}+j_{2i+2}+j_{6i+6}$  \\ \hline

$k=4$        & $k=4$                               & $k=4$      & $k=4$  \\ \hline
$g_{8i+2}$   & $4j_{4(8i+2)}$                      & $h_{8i+2}$ & $4j_{4(8i+2)}+2j_{2(8i+2)}+j_{8i+2}$  \\
$g_{8i+4}$   & $4j_{4(8i+4)}+2j_{2(2i+4)}$         & $h_{8i+4}$ & $4j_{4(8i+4)}+2j_{2(2i+4)}+2j_{2(8i+4)}$ \\
             &                                     &            & $+j_{8i+4}+j_{4i+2}$ \\  
$g_{8i+6}$   & $4j_{4(8i+6)}$                      & $h_{8i+6}$ & $4j_{4(8i+6)}+2j_{2(8i+6)}+j_{8i+6}$  \\
$g_{8i+8}$   & $4j_{4(8i+8)}+2j_{(8i+8)}$          & $h_{8i+8}$ & $4j_{4(8i+8)}+2j_{(8i+8)}+j_{2i+2}$  \\
             & $+j_{2i+2}$                         &            & $+2j_{2(8i+8)}+j_{8i+8}+j_{4i+4}$ \\  \hline

$k=5$        & $k=5$                               & $k=5$       & $k=5$  \\ \hline
$g_{10i+2}$  & $5j_{5(10i+2)}$                     & $h_{12i+2}$ & $g_{12i+2}+3j_{3(12i+2)}+2j_{2(12i+2)}+j_{12i+2}$    \\
$g_{10i+4}$  & $5j_{5(10i+4)}$                     & $h_{12i+4}$ & $g_{12i+4}+3j_{3(12i+4)}+2j_{2(12i+4)}$      \\
             &                                     &             & $+j_{12i+4}+j_{6i+2}$ \\
$g_{10i+6}$  & $5j_{5(10i+6)}$                     & $h_{12i+6}$ & $g_{12i+6}+3j_{3(12i+6)}+2j_{2(12i+6)}$   \\
             &                                     &             & $+j_{12i+6}+j_{4i+2}$  \\
$g_{10i+8}$  & $5j_{5(10i+8)}$                     & $h_{12i+8}$ & $g_{12i+8}+3j_{3(12i+8)}+2j_{2(12i+8)}$    \\
             &                                     &             & $+j_{12i+8}+j_{6i+4}$    \\
$g_{10i+10}$ & $5j_{5(10i+10)}+j_{2i+2}$           & $h_{12i+10}$& $g_{12i+10}+3j_{3(12i+10)}+2j_{2(12i+10)}+j_{12i+10}$  \\
             &                                     & $h_{12i+12}$& $g_{10i+12}+3j_{3(12i+12)}+2j_{2(12i+12)}+j_{12i+12}$ \\
             &                                     &             & $+j_{6i+6}+j_{4i+4}$\\
\hline
 \end{tabular}}\caption{Periodicity of the coefficients $g_{n}$ for $c=24\cdot k$ extremal partition functions $\mathcal{G}_{k}$, 
and for $h_{n}$ coefficients of characters of the extremal vertex operator algebras $\mathcal{H}_{k}$ 
in terms of coefficients the $j_{2n}$ of the modular function $j$}\label{tab-mon3}\end{table}
These results are somewhat reminiscent of Bott periodicity for the
stable homotopy of the classical groups. Here we are dealing with (the
equivalent of) increasing level algebras.

 To summarize, when $k=1$ it is known via standard Monster
Moonshine that the coefficients of $j$ decompose into Monster representations \cite{Borch}. 
The fact that all the higher $k$ coefficients also decompose into Monster representations indicates
that they have large symmetries containing the Monster and the fact that they have these symmetries may 
indicate that they are related to $24k$ dimensional lattices.


\section{Conclusions}

Using the techniques presented in \cite{Jankiewicz:2005rx}, one can construct a large class of conformal 
field theories with central charge that is a multiple of 24. We have demonstrated (or at least conjecture) 
the possibility of  a new realization of Monster moonshine. This is realized as a periodicity 
in a pattern of coefficients in $q$-expansions of the extremal partition functions.

\section*{Acknowledgments}
MJ thanks NSF and the QTS4 organizers for travel support.
This work was supported in part by U.S. DoE grant \#~DE-FG05-85ER40226.





\begin{thebibliography}{10}

\bibitem{Jankiewicz:2005rx}
  M.~Jankiewicz and T.~W.~Kephart,
  ``Transformations among large c conformal field theories,''
  arXiv:hep-th/0502190.
\bibitem{Dixon:1988qd}
L.~J.~Dixon, P.~H.~Ginsparg and J.~A.~Harvey,
Commun.\ Math.\ Phys.\  {\bf 119}, 221 (1988).
\bibitem{Apostol} Tom M. Apostol, ``Modular Functions and Dirichlet Series in Number Theory'', Graduate Texts in Mathematics, Springer-Verlag 1990.;
C. L. Siegel, Advanced Analytic Number Theory, Tata Institute of Fundamental Research, Bombay, 1980, pp. 249-268.
\bibitem{Conway}J.H. Conway, N.J.A. Sloane, ``Sphere Packings, Lattices and Groups'',  Springer-Verlag NY (1993).
\bibitem{Harvey:1988ur}J.~A.~Harvey and S.~G.~Naculich,
Nucl.\ Phys.\ B {\bf 305}, 417 (1988).
\bibitem{Dolan:1989kf}
  L.~Dolan, P.~Goddard and P.~Montague,
  Phys.\ Lett.\ B {\bf 236}, 165 (1990).;
  L.~Dolan, P.~Goddard and P.~Montague,  
  Nucl.\ Phys.\ B {\bf 338}, 529 (1990).;
L.~Dolan, P.~Goddard and P.~Montague, 
Commun.\ Math.\ Phys.\  {\bf 179}, 61 (1996)
P.~S.~Montague,
Nucl.\ Phys.\ B {\bf 428}, 233 (1994)
P.~S.~Montague,
Lett.\ Math.\ Phys.\  {\bf 44}, 105 (1998)
\bibitem{Borch}R.E. Borcherds, A.J.E. Ryba, Duke Math J. 83 (1996) no.2 435-459.;
 T.~Gannon, ``Monstrous moonshine and the classification of CFT,''
  arXiv:math.qa/9906167.







\end{thebibliography}
\end{document}